\documentclass[a4paper,10pt,twoside]{cpc-hepnp}
\usepackage{CJK,upgreek,fancyhdr}
\usepackage{multicol}
\usepackage{graphicx}
\usepackage{booktabs}
\usepackage{amssymb,bm,mathrsfs,bbm,amscd}
\usepackage[tbtags]{amsmath}
\usepackage{lastpage}
\usepackage{lineno}
\usepackage{hyperref}
\usepackage{ulem}

\begin{document}
\begin{CJK*}{GB}{gbsn}

\fancyhead[c]{\small Chinese Physics C~~~Vol. xx, No. x (201x) xxxxxx}
\fancyfoot[C]{\small 010201-\thepage}

\footnotetext[0]{Received 15 Apr. 2016}

\title{e$^+$/e$^-$ Discrimination in Liquid Scintillator and
        Its Usage to Suppress $^{8}$He/$^{9}$Li Backgrounds \thanks{
        Supported by National Natural Science Foundation of China (Grant
        No.~11575226,~11475197,~11205183). }}

            \author{%
                Ya-Ping CHENG (³ÌÑÅÆ») $^{1,2;1)}$\email{chengyp@mail.ihep.ac.cn}%
                    \quad Liang-Jian Wen (ÎÂÁ¼½£) $^{2}$\email{wenlj@mail.ihep.ac.cn}%
                    \quad Peng Zhang (ÕÅÅô) $^{2}$%
                    \quad Xing-Zhong Cao (²ÜÐËÖÒ ) $^{2}$%
            }
\maketitle

\address{%
    $^1$ University of Chinese Academy of Sciences, Beijing 100049,China\\
        $^2$ Institute of High Energy Physics, Chinese Academy of Sciences, Beijing 100049, China\\
}

\begin{abstract}
Reactor neutrino experiments build large-scale detector systems to detect
neutrinos. In liquid scintillator, a neutral bound state of a
positron and an electron, named positronium, can be formed. The spin
triplet state is called ortho-positronium (o-Ps). In this article, an experiment
is designed to measure the lifetime of o-Ps, giving a result of 3.1 ns. A PSD
parameter based on photon emission time distribution (PETD) was constructed to discriminate e$^+$/e$^-$.
Finally, the application of e$^+$/e$^-$ discrimination in the JUNO experiment is
shown. It helps
suppress $^{8}$He/$^{9}$Li backgrounds and improves the sensitivity
by 0.6 in $\chi^2$ analysis with an assumption of $\sigma$=1 ns PMT Transit Time
Spread, which will bring a smearing effect to the PETD.
\end{abstract}

\begin{keyword}
PSD, LS, ortho-positronium
\end{keyword}

\begin{pacs}
36.10.Dr, 29.40.Mc, 78.70.Bj
\end{pacs}

\footnotetext[0]{\hspace*{-3mm}\raisebox{0.3ex}{$\scriptstyle\copyright$}2013
Chinese Physical Society and the Institute of High Energy Physics
of the Chinese Academy of Sciences and the Institute
of Modern Physics of the Chinese Academy of Sciences and IOP Publishing Ltd}%

\begin{multicols}{2}

\section{Introduction}\label{sec:intro}
In most reactor neutrino experiments, electron antineutrinos
are detected via the inverse $\beta$-decay (IBD) reaction, $\bar{\nu_e} + p \rightarrow e^+ + n$.
The antineutrino signature is a coincidence between the prompt positron
signal and the microseconds-delayed neutron capture on the
target (proton or doped isotope). The most serious background is
the cosmogenic $^{8}$He/$^{9}$Li background, which can mimic the IBD
signature via a $\beta + n$ cascade decay. Thus, the capability of
positron-electron discrimination is extremely useful to reject the
$^{8}$He/$^{9}$Li background.

Prior to annihilation, the positron can form
a neutral bound state of a positron and an electron, namely positronium.
The formation threshold of positronium is 6.75 eV \cite{theorycal}.
Depending on the total spin angular momentum, positronium is classified
into the spin singlet para-positronium (p-Ps) state and the spin triplet
ortho-positronium (o-Ps) state.
In vacuum, the p-Ps state annihilates by emitting two $\gamma$ rays
of 511 keV with a mean lifetime of 125 ps, while the o-Ps state emits
three $\gamma$ rays with a mean lifetime of $\sim$140 ns.
In liquid scintillator (LS), the mean lifetime of o-Ps state is strongly
reduced to a few nanoseconds due to the interactions of o-Ps with the
surrounding medium~\cite{brex}. The o-Ps in different
LS solvents has slightly different lifetimes and formation
probabilities~\cite{addref}.

The delay of positron annihilation can induce distortion in the photon
emission time distribution (PETD) of the detected photons.
In addition, the two annihilation gamma rays will further induce
distortion to the PETD due to Compton scattering. The PETD difference
between positrons and electrons provides an opportunity for positron-electron
discrimination in LS detector and the consequent rejection of $^{8}$He/$^{9}$Li
background from IBD candidates.

In this paper, we study the potential mass hierarchy sensitivity improvement of
the Jiangmen Underground Neutrino Observatory (JUNO) (\cite{juno1,juno2}) by
positron-electron discrimination analysis.
JUNO is a 20 kton multi-purpose liquid scintillator
detector, with the primary goal to determine the
neutrino mass hierarchy by detecting reactor
antineutrinos. The JUNO central detector uses linear alkyl-benzene (LAB)-based LS and has excellent
energy resolution of $\sim$3\%/$\sqrt{E}$. The energy scale is about 1200 photoelectron/MeV and the PMT
photocathode coverage is about 78\%.
The outline of this article is as follows. In
Section~\hyperref[sec:pals]{2} a newly designed
apparatus for positron annihilation lifetime measurement is presented. In
Section~\hyperref[sec:dtda]{3} the measured o-Ps lifetime in LAB is reported.
Section~\hyperref[sec:sim]{4} describes a pulse shape discrimination (PSD) method
for positron-electron discrimination. Finally, the application of positron-electron discrimination to JUNO and the improvement on mass
hierarchy sensitivity is shown in Section~\hyperref[sec:psd]{5}.

\section{Positron lifetime measurement experimental apparatus}\label{sec:pals}
A Positron Annihilation Lifetime Spectroscopy (PALS) apparatus, developed in-house,
was used to measure the o-Ps formation fraction and its
lifetime. A schematic diagram of the experimental setup, and the
electronics and data acquisition (DAQ) scheme, are shown in Fig.~(\ref{fig:exp}).
The notations in Fig.~(\ref{fig:exp})b are as follows:
CFDD: constant fraction differential discriminator;
SCA: single channel analyzer;
TAC: time-to-amplitude converter;
MCA: multichannel amplitude analyzer.
\begin{center}
\includegraphics[width=7.5cm]{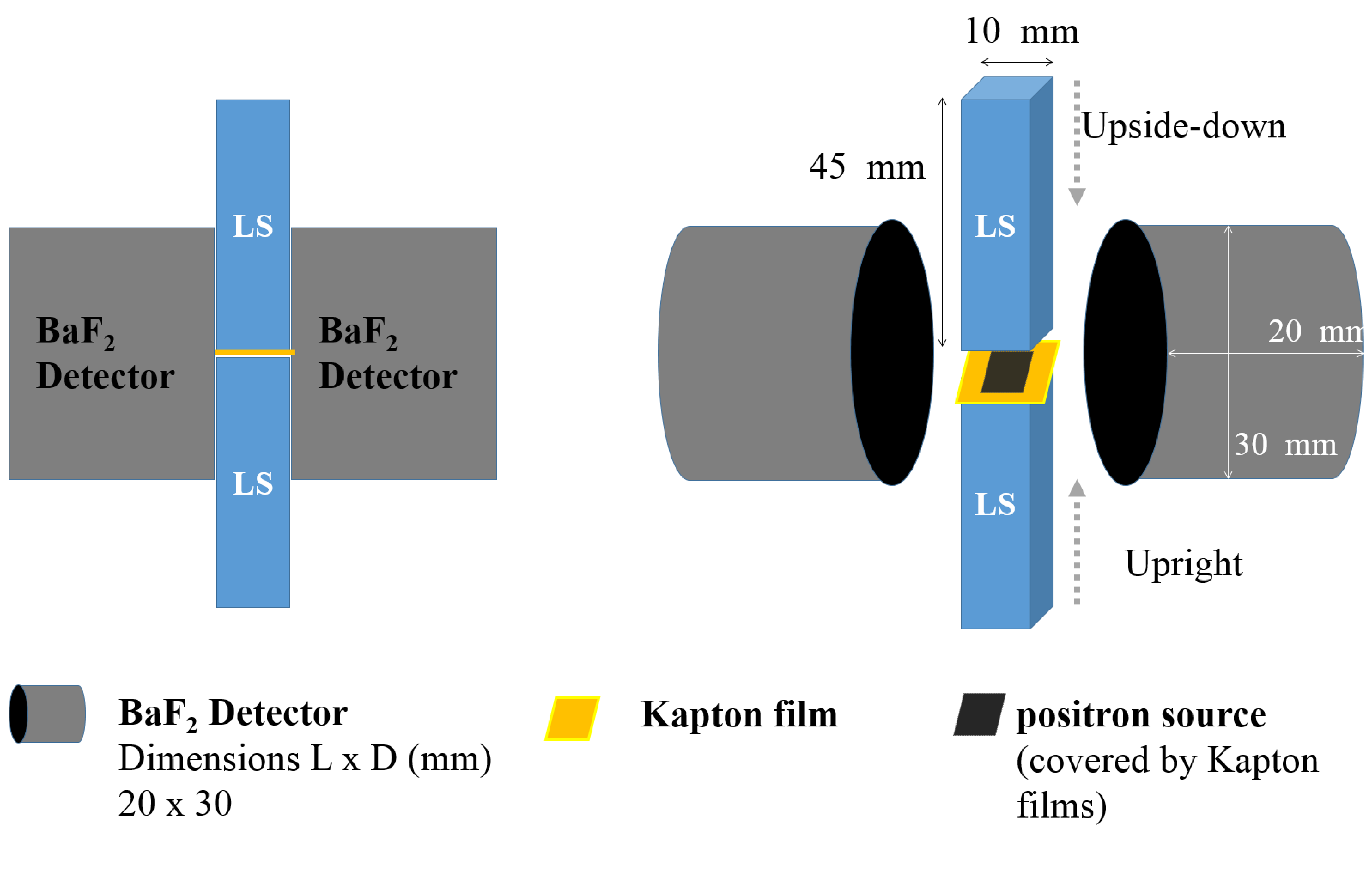}\\
\text{ (a) Schematic diagram of the experimental setup
}\\
\text{(Left: side view of the setup
\qquad \qquad }\\
\text{Right: dimensions of each constituent part)
}\\
\includegraphics[width=7.5cm]{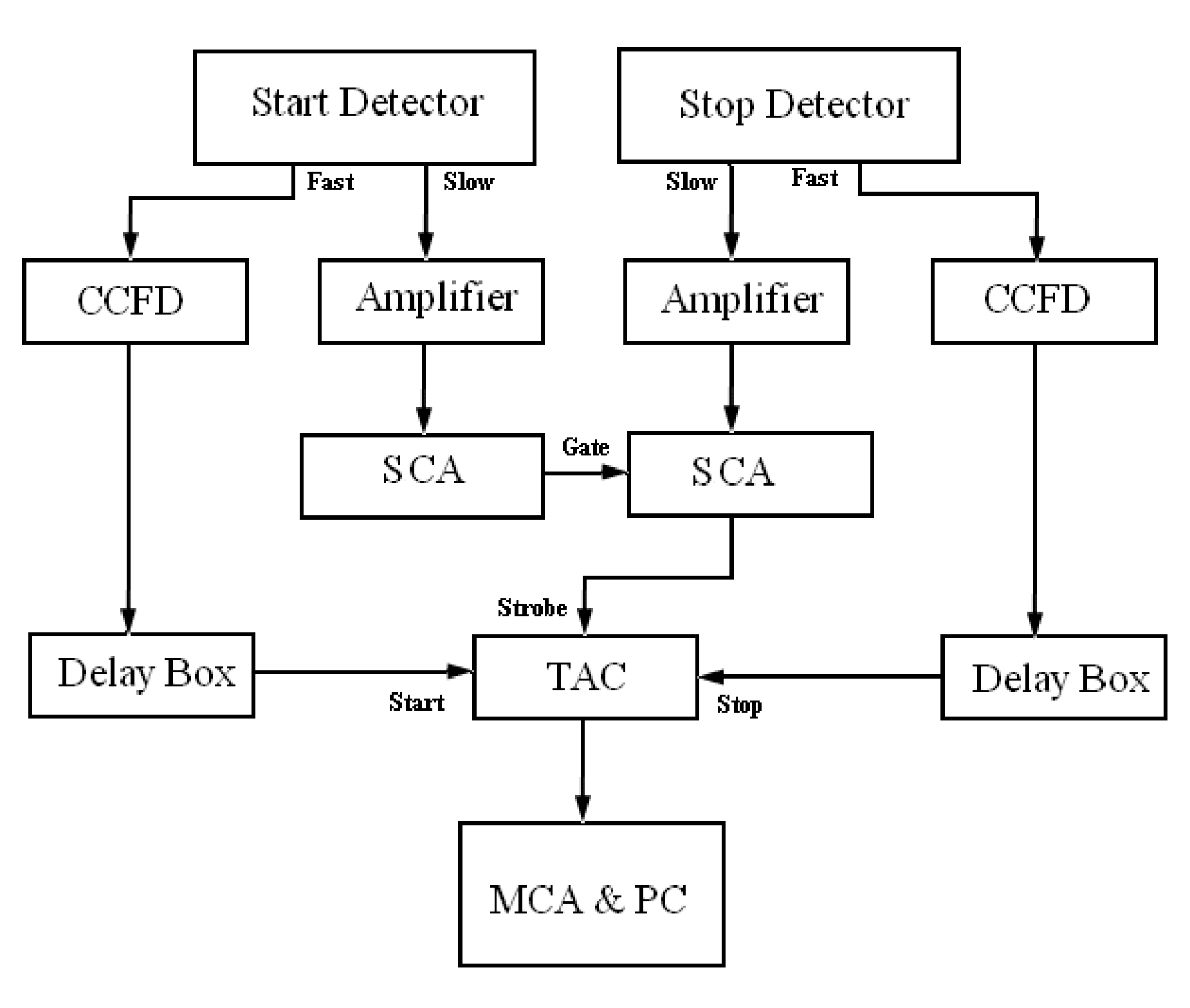}\\
\text{ (b) Electronics and data acquisition scheme }\\
\figcaption{\label{fig:exp} Experimental apparatus}
\end{center}

Typically the sample for the PALS is crystal or powder
pressed into thin slices. To measure the positron annihilation in
the LAB liquid, a special design was developed.
A thin positron source was made of a dried drop of liquid NaCl
encapsulated in two layers of 7.5 $\mu$m Kapton films, with $^{22}$Na serving as
the positron source. Over 90\% of radioactive $^{22}$Na
decays by emission of a positron to the excited state of $^{22}$Ne, then
immediately reaches the ground state by emission of a 1.274 MeV $\gamma$.
The kinetic energy spectrum of positron is a continuous distribution ranging from 0 to 545 keV
and the activity of the positron source was 15 $\mu$Ci.
The LS sample was placed in a disposable Brand cuvette \cite{brand}
made of acrylic material which is completely compatible with LS.
Each cuvette had a capacity of 4.5 mL (10 mm $\times$ 10 mm
$\times$ 45 mm)
 with no lid,
and was 1 mm thick,
sealed by a piece of thin Kapton film. The two cuvettes were stacked vertically
with the flat positron source placed in between the
two films.
The top cuvette was placed upside-down, and the bottom cuvette placed upright.
Two BaF$_{2}$ plastic scintillators (dimensions
L x D: 20 mm $\times$ 30 mm) coupled with 2-inch
PMTs (type: XP2020Q)
were covered by black film, placed at opposite sides of the cuvettes, and used to measure the 511 keV annihilation $\gamma$
and the 1.27 MeV $\gamma$.
The DAQ system was a fast-slow coincidence system.
The BaF$_2$ crystal emits both
fast and slow scintillation components. The decay time of fast and slow
components is $\sim$600 ps and $\sim$630 ns respectively, and their intensities
are $\sim$15\% and $\sim$85\% respectively.
The fast coincidence channel was for time measurement. The
fast component signals from the BaF$_{2}$ detectors,
which served as the timing signal, were passed to
the input of a constant fraction differential discriminator (CFDD).
The slow signals from the BaF$_{2}$ detectors were fed to the slow coincidence
channel.
Two single channel analyzers (SCAs) were used to
select $\gamma$s over the energy range of 1.27 MeV and 511 keV.
The 1.27 MeV $\gamma$ from $^{22}$Na decay served as the start signal, and
the 511 keV $\gamma$ from positron annihilation served as the stop signal.
The appropriate $\gamma$s
generated the strobe signal for the time-to-amplitude
converter (TAC). The analog pulse proportional to the time interval
between the start and stop pulses was fed to the multichannel
amplitude analyzer (MCA) and converted into a digital signal,
then stored on data disk for later analysis.

\section{Data taking and data analysis}\label{sec:dtda}
Oxygen dissolved in liquid scintillator can decrease
the positron annihilation lifetime. To remove oxygen in the test sample,
we bubbled the sample with high purity nitrogen
for about 40 minutes.
The flow rate of nitrogen was carefully controlled by
a pressure reducing valve. The time interval between the 511 keV $\gamma$ from
positron annihilation and the 1.27 MeV $\gamma$ which accompanies the sodium
$\beta^+$
decay was taken as the lifetime of the positron. The time resolution of the
system was 190 ps (FWHM)
calibrated by the two
instantaneous $\gamma$ rays of a cobalt-60 source.

The o-Ps lifetime was extracted from the measured lifetime spectrum
using the RooFit~\cite{roofit} tool.
The positron annihilation lifetime in LS can be described as the sum of two
exponential components. The short-lived p-Ps and directly annihilating positron
are indistinguishable, therefore they are merged into one component,
denoted as
$\tau_{0}$ in Eq.~(\ref{lifefit}). The other
component represents o-Ps, shown as $\tau_{1}$ in Eq.~(\ref{lifefit}). The
formation probability of o-Ps is represented by $w$ in
Eq.~(\ref{lifefit}). Positron annihilation in the substrate material also
contributes to the measured lifetime spectrum. This
can be described by two exponential components. One originates from the Kapton,
denoted as $\tau_{0s}$. The other component is generated from contaminants in
the positron source, denoted as $\tau_{1s}$. Similarly, the relative weight of
these
two components is shown as $w_s$. The percentage of positron
annihilation in the source is evaluated by $I_s$ in Eq.~(\ref{lifefit}).
The system instrumental time resolution is assumed as a Gaussian distribution.
Using $\sigma$ and $T$ to represent the instrumental time resolution and time latency,
and $t^{\prime}$ as the convolution operation sign, the final fitting function is shown as Eq.~(\ref{lifefit}).

\begin{eqnarray}
\label{lifefit}
F(t)=&\int_{0}^{t}\frac{N}{\sqrt{2\pi}\sigma}e^{-\frac{t-T-t^{\prime}}{2\sigma^2}}
\Big\{
     (1-I_s)\times (
    \frac{1-\omega}{\tau_{0}}e^{-\frac{t^{\prime}}{\tau_{0}}}+
    \frac{\omega}{\tau_{1}}e^{-\frac{t^{\prime}}{\tau_{1}}}
    )
\nonumber \\
     &+I_s \times (
    \frac{1-\omega_s}{\tau_{0s}}e^{-\frac{t^{\prime}}{\tau_{0s}}}+
    \frac{\omega_s}{\tau_{1s}}e^{-\frac{t^{\prime}}{\tau_{1s}}}
    )
\Big\}
dt^{\prime}
\end{eqnarray}
The Life Time 9.0 program~\cite{lt9} (LT9), based on the least squares
fitting method, was used as a crosscheck.
Their difference was taken as the error in the measured positron annihilation lifetime.
In this measurement, optical photons were
shielded by the black film and only $\gamma$s can propagate into the BaF$_{2}$
detector, thus in the measured lifetime spectrum there
is no contribution from the LS sample's scintillation light.

A standard sample made from nickel was measured first, and the effects of the source
substrate were obtained by fitting the measured lifetime spectrum of nickel. The
positron annihilation lifetime in nickel is described by
one exponential component and known to be about 108 ps~\cite{thecal}. The
results from LT9 and RooFit were 117 ps and 112 ps respectively. The source effects
are generated from Kapton and other impurities.
Positron annihilation lifetime in
impurities
($\tau_{1s}$) and w$_s$
were fixed by the value from nickel fitting.
$\tau_{0s}$, the positron lifetime in Kapton, was treated as a
floating parameter.
In the fitting for the nickel sample,
$\tau_{0s}$ was 385 ps.
In the fitting for the LAB sample,
$\tau_{0s}$
was
about 382 ps.
The positron lifetime in impurities $\tau_{1s}$ and its intensity w$_s$
were 1.095 ns and 11.88\% respectively.
I$_s$, the percentage of positrons lost in the positron source (Kapton or
other impurities), was about 50.34\%.
$\tau_{0}$, the lifetime of p-Ps or directly annihilating positrons, was 150 ps.
The
positron annihilation lifetime spectrum in LAB is shown in Fig.~(\ref{lab}).
The top panel shows the measured and fitted
lifetime spectrum.
The bottom panel shows the residual between data and fit normalized to
the statistical Poisson error of fit counts. From the unfolding result,
the o-Ps lifetime is about 3.10 $\pm$ 0.07 ns  and its
formation probability is about (43.7 $\pm$ 1.2)\%.
   \begin{center}
   \includegraphics[width=8.0cm]{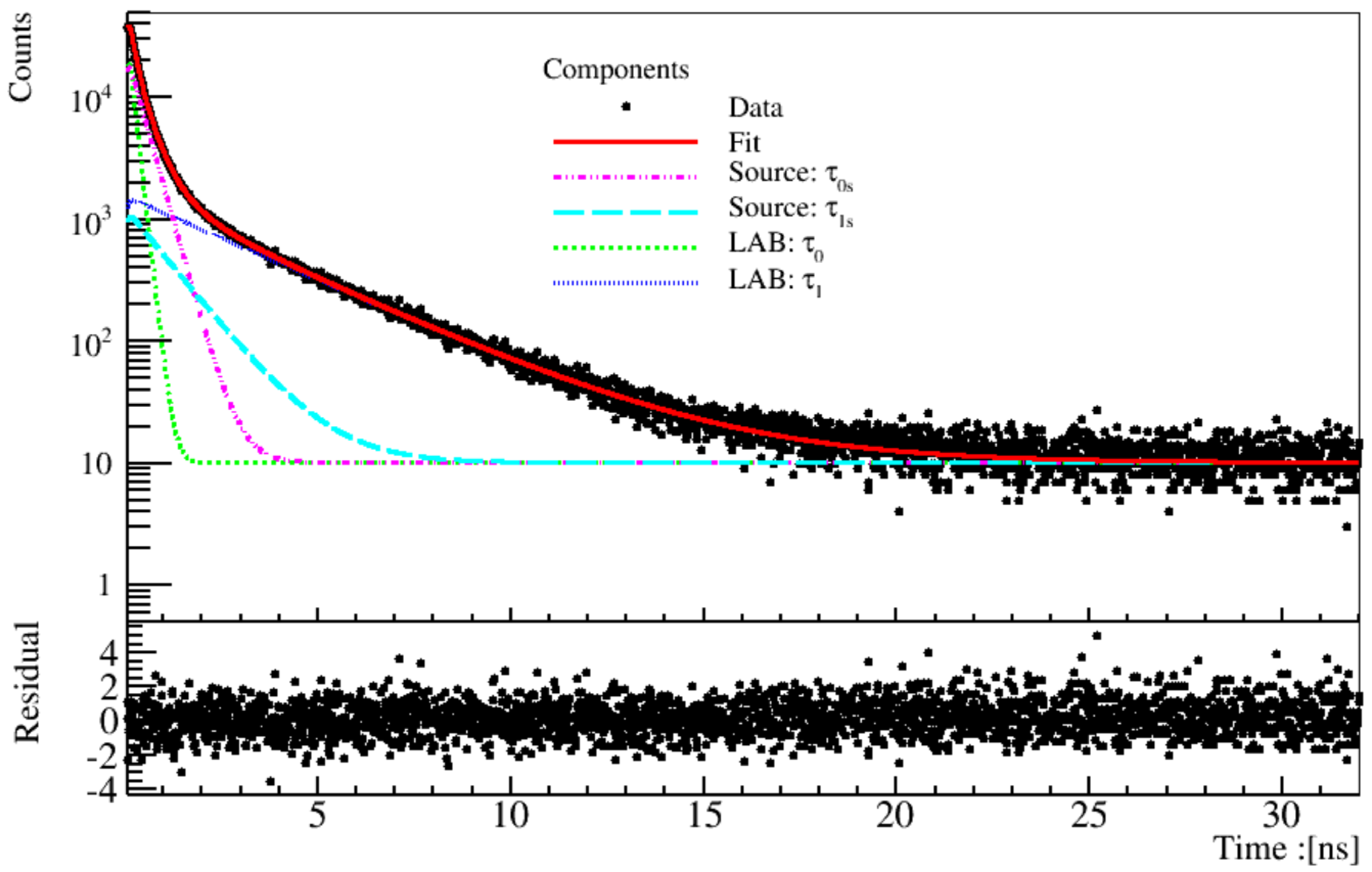}
   \figcaption{\label{lab} The measured and fitted lifetime spectra of positron annihilation in LAB. }
   \end{center}

\section{Simulation and PSD performance}\label{sec:sim}
Based on the measurement result described in the previous section, we studied the
e$^+$/e$^-$ discrimination power by simulation.
The simulation was done using the JUNO simulation framework SNiPER
\cite{citesniper}, in which the positronium production process is not
included in the physics list.
Therefore, for simplicity, a tag on photons
generated directly from the positron annihilation process was created. We did a
sampling based on o-Ps formation probability 43.7\% and lifetime
3.1 ns. The sampling result served as the delay to
photon emission time caused by o-Ps for those tagged photons.
Under such assumptions, taking positrons and electrons with deposited energy 2.5
MeV as an example, we got the averaged time profile for positron and electron
as Fig.~(\ref{waveform}). The time profile was determined by measuring the
arrival times of
both early and late hits
on each PMT in one event.

   \begin{center}
   \includegraphics[width=7.5cm]{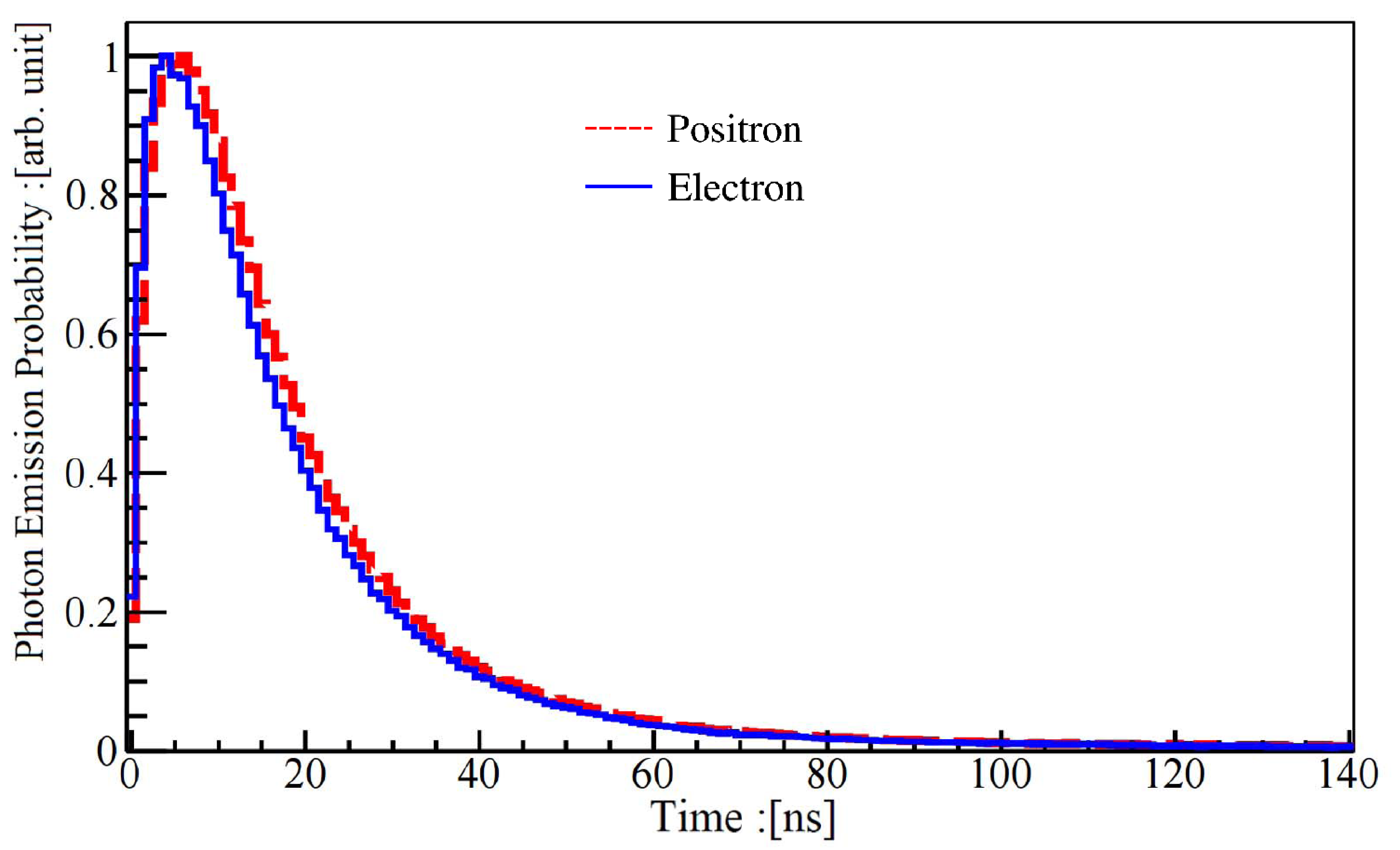}
   \figcaption{\label{waveform} Photon emission time profile for 2.5 MeV positrons and
   electrons as an example}
   \end{center}

To exploit the difference, we created PSD
parameters.
Two PSD parameters were constructed to do e$^+$ and e$^-$ discrimination.
One parameter is the tail to total ratio in the PETD. The start of the tail was chosen to be 10 ns and the total time
window was chosen to be 50 ns. The other parameter is the optimum Gatti parameter~\cite{psd_parameter}
defined as Eq.~(\ref{Gatti}). Here $r_i(t_n)$ denotes the photon emission
probability in the time interval $t_{n-1}$ to $t_{n}$ for the unknown
particle. The weight $w(t_{n})$ was calculated using e$^+$ photon
emission probability ($r_1(t_n)$) and e$^-$ photon emission probability
($r_2(t_n)$). The performance of the tail to total ratio parameter was not as
good as the Gatti parameter, so it was not used.

\begin{eqnarray}
\label{Gatti}
G=\sum_{n}{r_{i}(t_{n})w(t_{n})}\quad\quad\quad\quad\quad\\
r_{i}(t_{n})=\int_{t_{n-1}}^{t_{n}}P_{i}(t)dx \quad\quad
w(t_{n})=\frac{r_{1}(t_{n})-r_{2}(t_{n})}{r_{1}(t_{n})+r_{2}(t_{n})} \nonumber
\end{eqnarray}

Besides these two PSD parameters, an artificial neutral
network method was also investigated. The multilayer perceptron (MLP) algorithm
from the ROOT Toolkit for Multivariate Data Analysis~\cite{tmva}
was used.
Such an algorithm yields an output qualifier indicating the particle type.
The input
to the algorithm comprises 16 variables, defined as follow. We build the
cumulative distribution function of
the PETDs, and we set 16 thresholds equally distributed between 0.03 and 0.33.
The time values at which the
threshold crossings occur serve as input variables. The total number and the
values of the thresholds were
optimised to sample the PETD range with the most discrimination power.
The training and test samples are every other two entries in the ROOT tree file.

A figure of merit (FOM) parameter was used to
evaluate the PSD parameter discrepancy of positron and electron. FOM
was defined as the peak distance divided by the
summation of FWHM of each particle. We applied the Gatti parameter
and MLP algorithm on positron and electron Monte Carlo samples,
and found that the Gatti parameter and MLP had similar performance.
An example is shown in Fig.~(\ref{psdoutput}). The electron and positron
energy was
4.5 MeV, the Gatti FOM was 0.265 and the MLP FOM was 0.266.
The following analysis
is based on the Gatti discrimination parameter.

\begin{center}
\includegraphics[width=7.0cm]{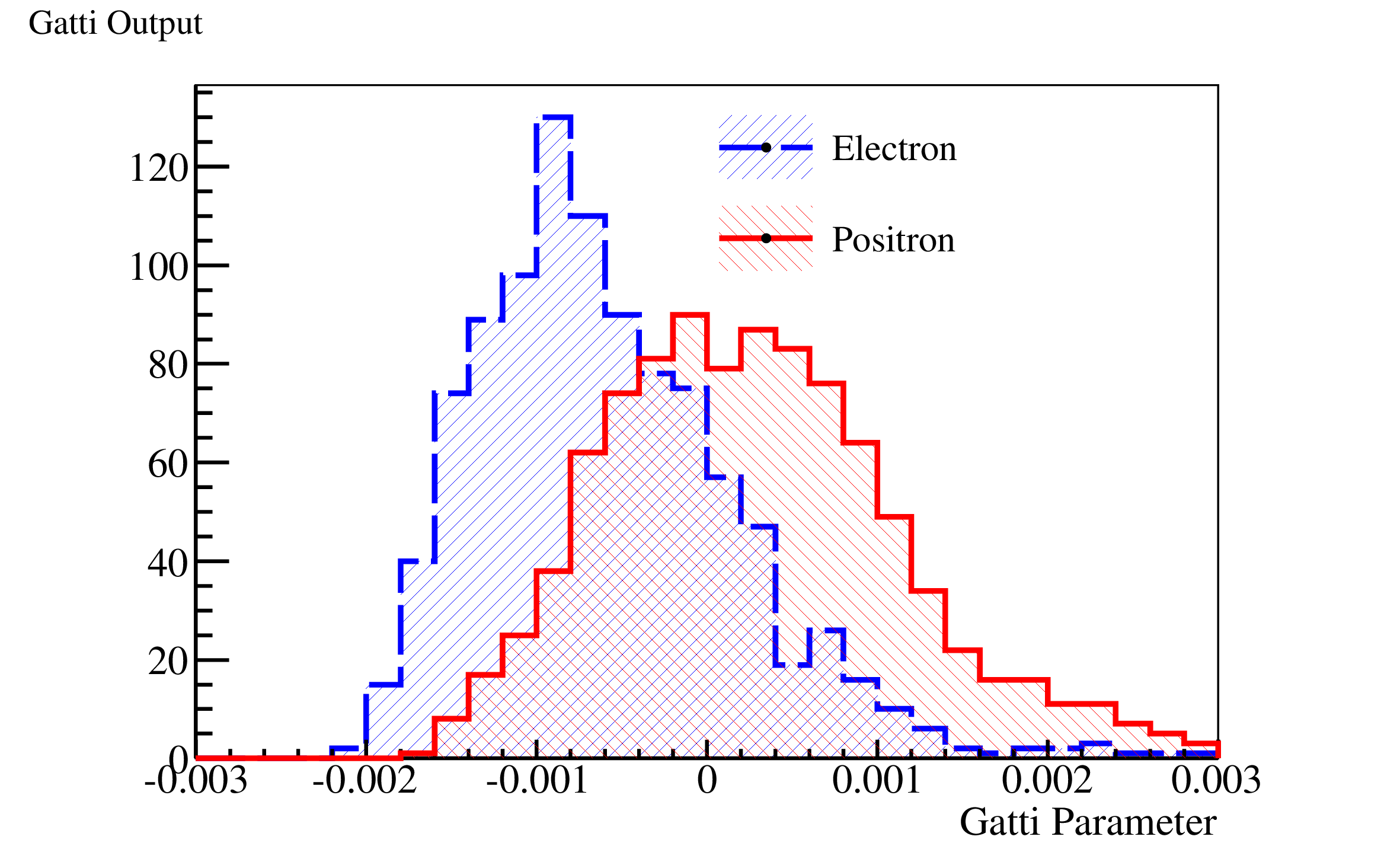}\\
\text{ (a) PSD output by Gatti parameter } \\
\includegraphics[width=7.0cm]{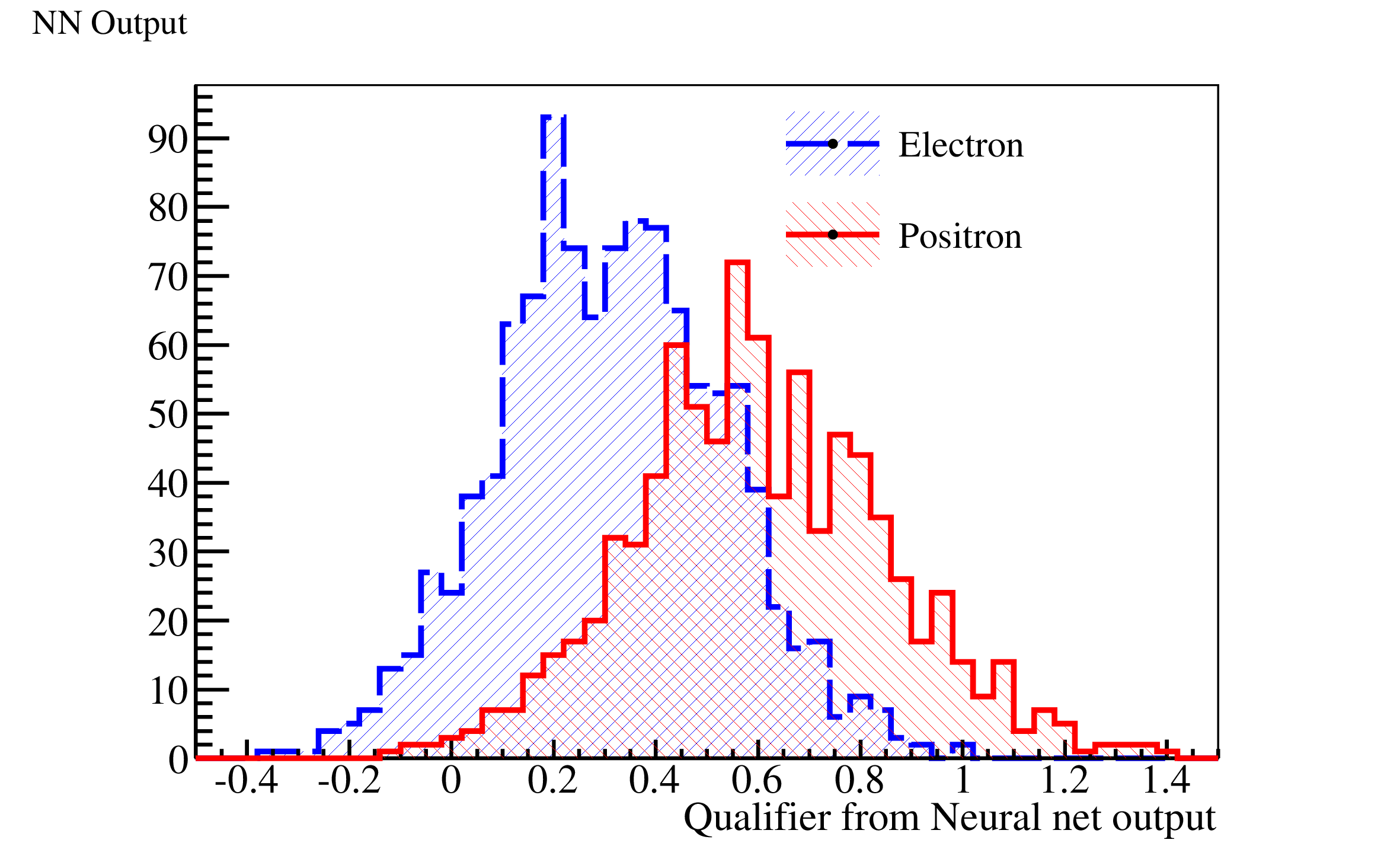}\\
\text{ (b) PSD output by MLP neural network }\\
\figcaption{\label{psdoutput} PSD performance of Gatti parameter and MLP
}
\end{center}

In consideration of the following studies, e$^+$/e$^-$ discrimination power at different
energies was studied. The energy range extended from 1.0 MeV ( the
minimum deposited energy of the positron) to 9.0 MeV.
In this study $\sigma=1$ ns PMT Transit Time Spread (TTS) was added to the PETD of
positron and electron.
The smearing effect on PETD from vertex resolution was much less severe. A 10
cm vertex reconstruction resolution was applied.
The e$^+$/e$^-$ discrimination at different energies is shown in Fig.~(\ref{fomeff}).

\begin{center}
\includegraphics[width=7.5cm]{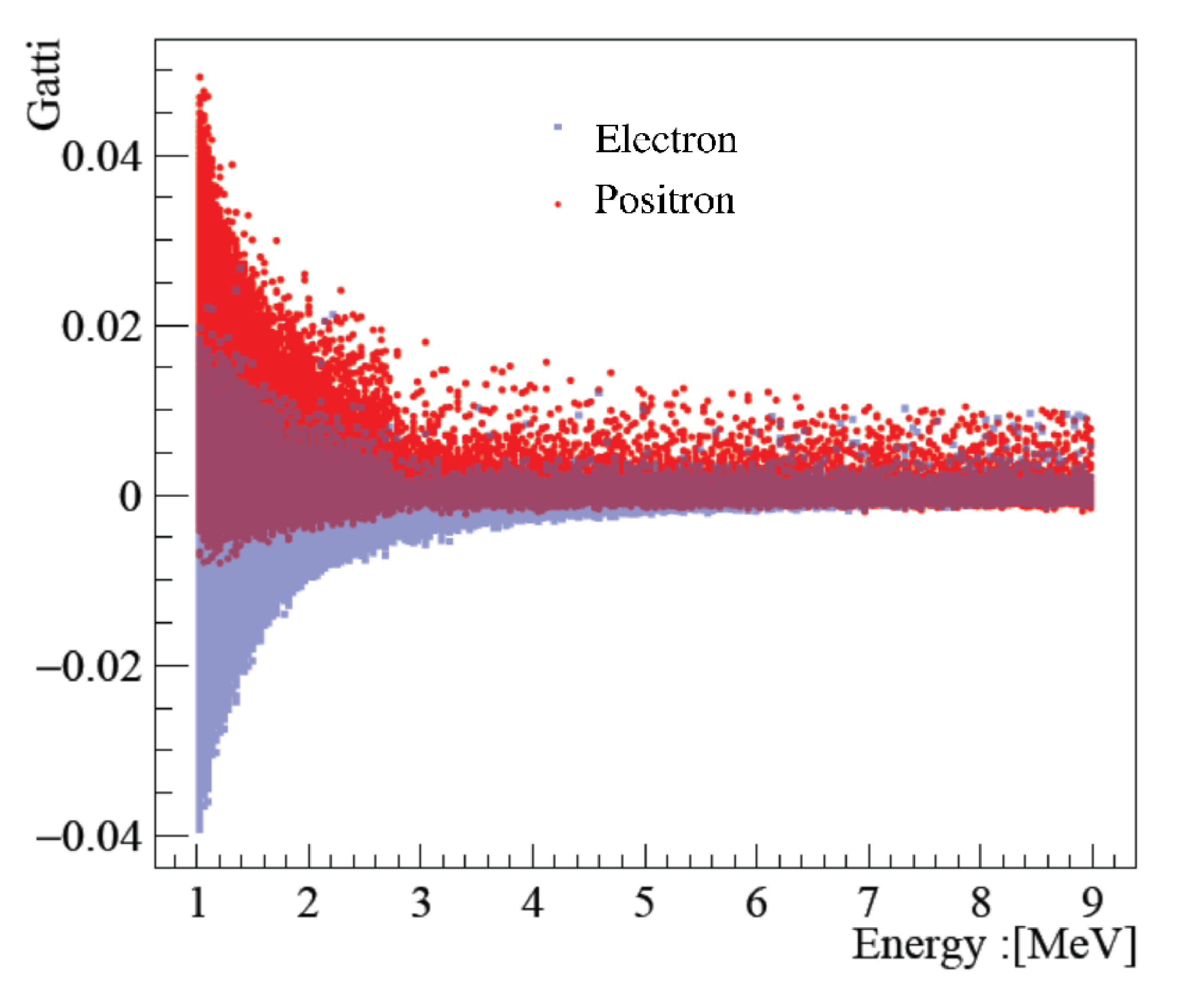}
\figcaption{\label{fomeff}  Gatti parameter distribution for e$^+$ and e$^-$
at different energies}
\end{center}

\section{PSD application in the JUNO experiment}\label{sec:psd}
$^{8}$He and $^{9}$Li can mimic IBD interactions.
The prompt signal includes the contributions from
electron, $\alpha$, and the recoil energy of the neutron.
The delayed signal is the neutron capture signal.
Since the electron contributes $\sim$90\% of the prompt signal's
visible energy and the PETD changes less than 0.5\%
compared with pure electron signals, in this study, the electron was taken as the
prompt signal.
Since the $^{8}$He and $^{9}$Li isotopes are mainly produced by cosmic
muons going through the detector, a veto on the detector volume within a time
window since the last muon can be applied to reject the $^{8}$He/$^{9}$Li background.
This veto strategy was studied in \cite{juno2}. For example, a veto cut to reject the detector volume within a cylinder with distance from muon track less than 3 meters in a 1.2~s time window since the last muon can reduce this background from 71/day to 1.6/day.
However, the IBD event rate is also reduced by 17\% due to the dead time caused by the veto.
The PSD in this study can be used as another method to reject $^{8}$He/$^{9}$Li background.
With the help of the PSD method, the veto cut can be loosened to reduce the
dead time without sacrificing signal to background ratio.
To evaluate the improvement of the mass hierarchy sensitivity by optimizing the
PSD and veto conditions, a $\chi^2$ was defined as Eq.~(\ref{chi2}),
where $i$ is the bin index of the spectra, $M^{i}$ is the simulated spectrum including IBD signal ($S^i$) and background
($B^i_b$), and $F^{i}$ is the fit spectrum with the oscillation parameters to be fitted.
${\varepsilon_R,\varepsilon_{r}, \varepsilon_B}$ are nuisance parameters
corresponding to the reactor flux and detector efficiency normalization factor,
reactor uncorrelated uncertainty ($\sigma_r$), and background rate uncertainty ($\sigma_B$). The spectrum shape uncertainties due to the IBD signal and background are
included by introducing $\sigma_{b2b}$ and $\sigma_b$.
    \begin{eqnarray}
    \label{chi2}
    \chi^{2} =
    \sum_i^{N_{bin}}  \frac{(M^{i}-F^{i})^2}{M^{i}+(\sigma_{b2b}S^i)^{2}+\sum_{Bkg}(\sigma_{b}B^i_b)^2}\nonumber \\
        +\sum_{Bkg}(\frac{\varepsilon_B}{\sigma_B})^2
        +\sum_r(\frac{\varepsilon_r}{\sigma_r})^2 \quad \quad \quad \quad \\
    F^i =
S^i(1+\varepsilon_R+\sum_{r}w_r\varepsilon_{r})+\sum_{Bkg}B^i_b(1+\varepsilon_B)
    \nonumber
    \end{eqnarray}
The simulated spectrum $M^{i}$ is calculated assuming either normal hierarchy (NH) or
inverted hierarchy (IH) without statistical fluctuations. When the fit spectrum uses
the same assumption, the minimization of the $\chi^2$ over the oscillation parameters
and nuisance parameters yields 0. While assuming the opposite mass hierarchy in the fit
spectrum, we obtain $\chi^2_{min}$. The sensitivity of the mass hierarchy can be expressed
as $\Delta \chi^2 = \chi^2_{min} - 0$.

The neutrino spectra after applying PSD analysis
served as the input for the ${\chi}^2$ analysis.
The PSD cut efficiency at each energy bin of the IBD prompt and
background spectrum was calculated by using mono-energetic MC e$^-$/e$^+$
samples.
Since we use the neutrino spectrum in the $\chi^2$ analysis, we need to do a
transformation between the prompt energy and neutrino energy to get the
corresponding efficiency at the neutrino spectrum bin. The conversion
was done according to equation 11 in \cite{ep2enu}. The PSD
efficiencies were applied on $M^{i}$, $S^i$ and the specific $B^i_b$ corresponding to
$^{8}$He/$^{9}$Li in Eq.~(\ref{chi2}). The PSD cut efficiency errors were taken as
uncorrelated systematic errors and were added in quadrature to $\sigma_{b2b}$ and
$\sigma_{b}$ in Eq.~(\ref{chi2}).

In the $\Delta \chi^2$ calculation, we scanned the PSD cut and veto conditions.
The tested veto schema included a combination of volume veto cylinder radius 1.6 m, 2 m,
3 m and veto time 0.7 s, 1 s, and 1.3 s.
One optimized muon veto scheme that gave good
$\chi^2$ was to veto the detector volume within a cylinder with distance from
the muon track less than 2 m in a 1.0 s time window.
We can improve the $\chi^2$ from
the original 10.60 without PSD analysis to
11.17 after PSD analysis.
The signal and background Gatti parameter distributions are
shown in Fig.~(\ref{mixGattiandpsdcutsel}).
\begin{center}
\includegraphics[width=7.5cm]{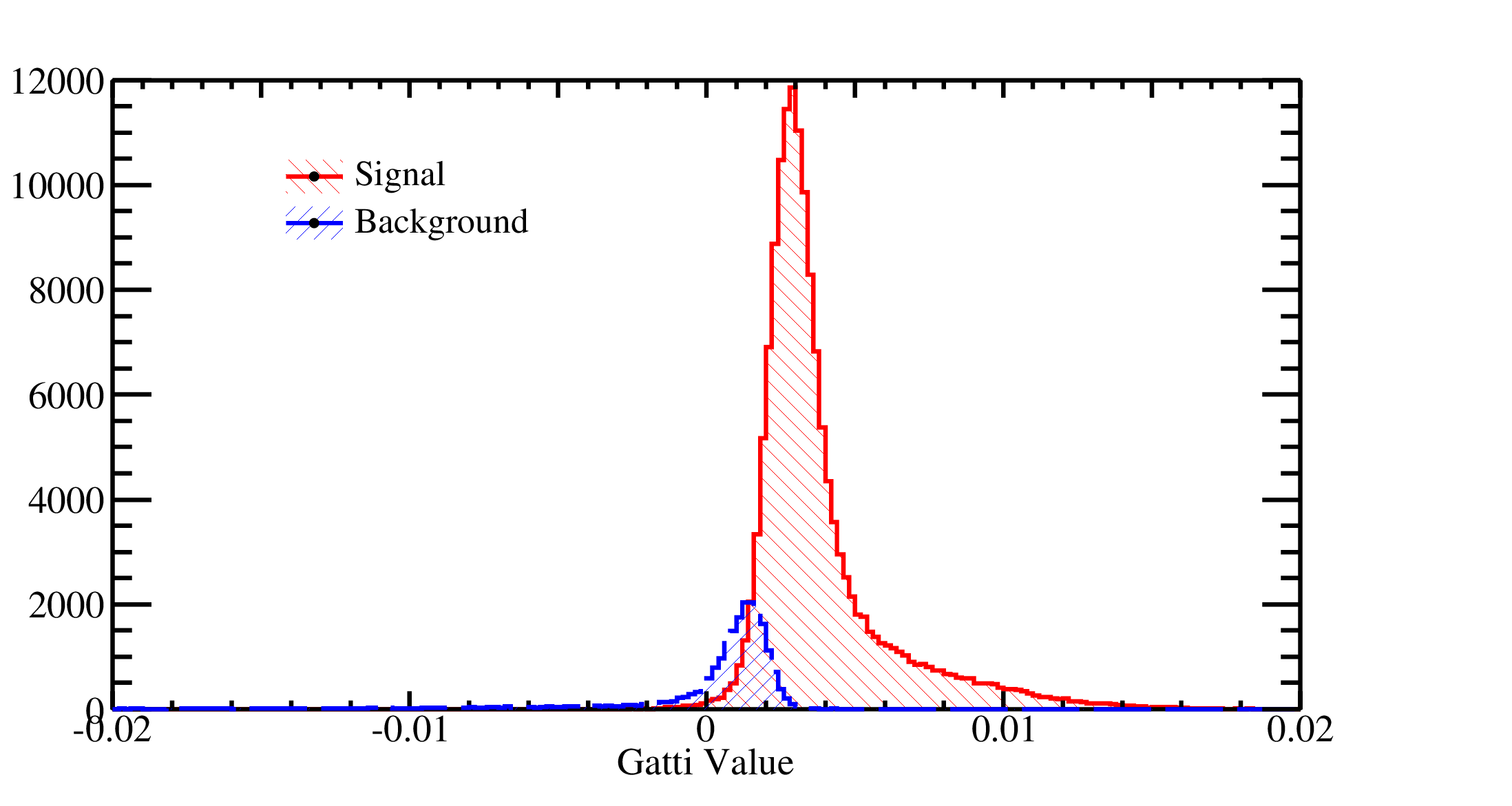}
\figcaption{\label{mixGattiandpsdcutsel} Mixed Gatti parameter distribution}
\end{center}
Figure~(\ref{finaleff}) shows the signal and background efficiency under the
optimized Gatti parameter cut ( -0.001 ).
In each energy bin, 4000 events were used and the statistical uncertainty
was calculated by carrying out the Bayesian approach, treating the number of
passing events as a binomially distributed variable, with uniform prior
probability assumption for the cut efficiency, shown as the error bars in
Fig.~(\ref{finaleff}).
The errors in Fig.~(\ref{finaleff}) are enlarged 10 times to be seen clearly.
In future, the PSD efficiency and uncertainties can be
derived from a data-driven analysis.

\begin{center}
\includegraphics[width=7.9cm]{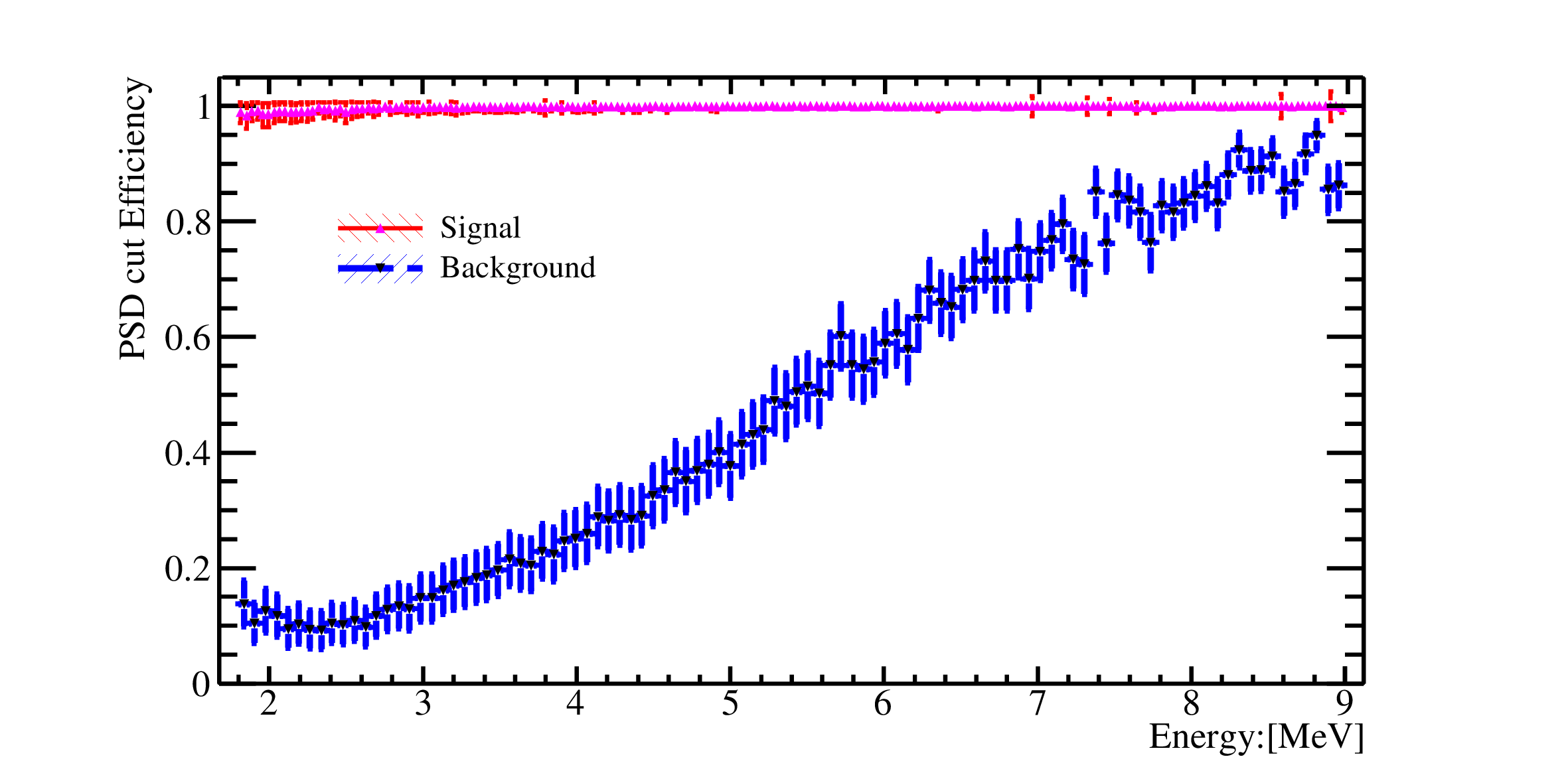}
\figcaption{\label{finaleff}   Signal and background efficiency under the
           optimized Gatti parameter cut, with efficiency error enlarged 10 times}
\end{center}

\section{Conclusion}\label{sec:cons}
We have measured the lifetime and formation probability of ortho-positronium.
In oxygen-free LAB, the o-Ps lifetime is 3.10 $\pm$ 0.07 ns and its formation probability is
(43.7 $\pm$ 1.2)\%.
The backgrounds in this design scheme were
calibrated by testing standard samples.

We performed simulations under the framework SNiPER. We assumed the PMT TTS was 1 ns
and the energy scale was about 1200 photoelectron/MeV, and we applied e$^+$/e$^-$ PSD to
reduce $^{8}$He/$^{9}$Li backgrounds. From the MH sensitivity study, by
loosening the muon veto cut and instead using
PSD analysis to reject the $^{8}$He/$^{9}$Li backgrounds, $\chi^2$ improvements
can be made. For example, vetoing the detector volume within 2 meters in a
1.0~s time window can improve the final $\chi^2$ from 10.60 (w/o PSD) to 11.06 (w/PSD).
\\

\acknowledgments{The authors would like to thank Dr. Liang Zhan for the helpful $\chi^2$
discussions, Dr. Miao He for his useful comments and Dr. Marco Grassi for his kindly language
editing help.}

\end{multicols}

\vspace{15mm}

\begin{multicols}{2}

\subsection*{Appendix A}

\begin{small}

\noindent{\bf Positron annihilation lifetime in nickel}

Positron annihilation lifetime in nickel was described by one exponential
component $\tau_{ni}$, instead of both $\tau_{0}$ and $\tau_{1}$ as shown
in
Eq.~(\ref{lifefit}). The measured and fitted spectra are
shown in Fig.~(\ref{nifit}). The top panel shows the measured and fitted
lifetime spectrum, with the bottom panel showing the
residual.

\begin{center}
\includegraphics[width=7.6cm]{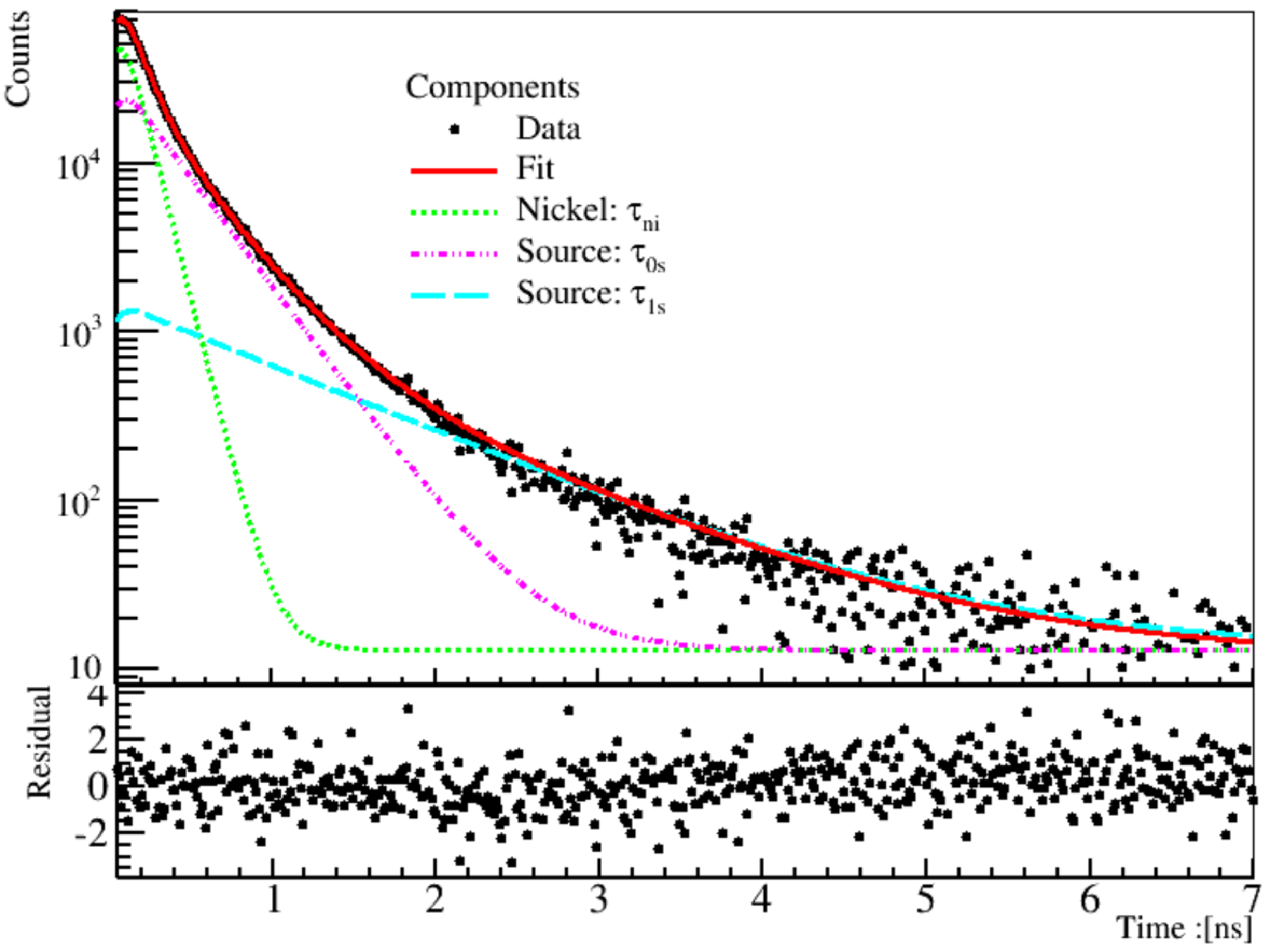}
\figcaption{\label{nifit}
The measured and fitted lifetime spectra of positron annihilation in
nickel}
\end{center}

\end{small}
\end{multicols}

\vspace{-1mm}
\centerline{\rule{80mm}{0.1pt}}
\vspace{2mm}

\begin{multicols}{2}

\end{multicols}

\clearpage
\end{CJK*}
\end{document}